\documentstyle[emulateapj]{article}

\lefthead{Pustilnik et al.}
\righthead{VLA HI Line Observations of SBS~0335--052}

\begin{document}

\title{VLA HI Line Observations of the Extremely Metal--Poor Blue Compact
Dwarf Galaxy SBS~0335--052}

\author{Simon A.\ Pustilnik}
\affil{Special Astrophysical Observatory, Russian Academy of Sciences,
 and Isaac Newton Institute of Chile, SAO Branch,
Nizhnij Arkhyz, 369167, Russia}
\authoremail{sap@sao.ru}

\author{Elias Brinks}
\affil{Departamento de Astronom\'{\i}a, Universidad de Guanajuato, Gto 36000
M\'exico}
\authoremail{ebrinks@astro.ugto.mx}

\author{Trinh X.Thuan} 
\affil{Astronomy Department, University of Virginia, Charlottesville VA 22903}
\authoremail{txt@starburst.astro.virginia.edu}

\author{Valentin A. Lipovetsky\altaffilmark{1}}
\affil{Special Astrophysical Observatory, Russian Academy of Sciences, Nizhnij Arkhyz, 369167, Russia}

\and

\author{Yuri I. Izotov}
\affil{Main Astronomical Observatory, National Academy of Sciences of Ukraine,
03680, Kyiv, Ukraine}
\authoremail{izotov@mao.kiev.ua}

\altaffiltext{1}{Deceased 1996, September 22.}

\begin{abstract}

We present the results of HI mapping with the NRAO\footnote{The
National Radio Astronomy Observatory is operated by Associated
Universities, Inc., under contract with the National Science
Foundation.} VLA of one of the most metal--deficient blue compact dwarf (BCD)
galaxies known, SBS~0335--052, with an oxygen abundance of only
1/40 that 
of the Sun.  We study the structure and dynamics of the neutral gas in
this chemically young object with a spatial resolution of 20\farcs5
$\times$ 15\arcsec\ ($\sim$ 5.4 $\times$ 3.9 kpc at an assumed distance of
54.3\,Mpc), a sensitivity at the 2$\sigma$ detection level of $\sim$2.0
K or $ 7.5 \times 10^{19}$ cm$^{-2}$ and a velocity resolution
of 21.2 km\,s$^{-1}$. We detected a large HI complex associated
with this object with an overall size of about 66 by 22 kpc and
elongated in the East--West direction.  There are two prominent,
slightly resolved peaks visible in the integrated HI map, separated in
the East--West direction by 22 kpc (84\arcsec).  The eastern peak
is nearly 
coincident with the position of the optical galaxy SBS~0335--052.  The
western peak is about a factor of 1.3 brighter in the HI line and is
identified with a faint blue compact dwarf galaxy, SBS 0335--052W,
with $m_B =$ 19.4, and a metallicity close to the lowest values known
for BCDs, about 1/50 that of the Sun.
The radial
velocities of both systems are similar, suggesting that the two BCDs
SBS~0335--052 and SBS~0335--052W constitute a pair of dwarf galaxies
embedded in a common HI envelope. Alternatively, the BCDs can be the 
nuclei of two distinct interacting primordial HI clouds.

The estimated total dynamical
mass, assuming the BCDs form a bound system, is larger than $\sim 6
\times 10^{9}$\,$M_{\odot}$. This is to be compared to a total gaseous
mass $M_{gas} = 2.1 \times 10^{9}$ $M_{\odot}$, and a total stellar
mass $M_{star}$   $\leq 10^{8}$ $M_{\odot}$. Hence, the mass of
the SBS~0335--052 system is dominated by dark matter.  
Because of the disturbed HI velocity field and the presence of what might be
tidal tails at either end of the system, we favor the hypothesis of tidal
triggering of the star formation in this system. It can be
due to either the nearby giant galaxy NGC~1376 or
the mutual gravitational interaction of the two HI clouds.

\end{abstract}


\section{Introduction}

Galaxy formation remains a key issue in observational 
cosmology. While much progress has been made in
finding large populations of galaxies at high ($z > 3$) redshifts 
(e.g. Steidel et al. 1996; Dey et al. 1998), 
truly young galaxies in the process of forming, defined as objects which
are experiencing their first epoch of star formation, remain
elusive. The spectra of those distant galaxies
generally indicate the presence of a substantial amount of heavy
elements, implying previous star formation and metal enrichment. 

We adopt here a different approach for studying truly young
galaxies.  Instead of searching for high redshift objects, we look
for nearby young dwarf galaxies which are possibly undergoing their
first burst of star formation.  As pointed out more than a quarter of
a century ago by Sargent \& Searle (1970) and Searle \& Sargent
(1972), some low--mass galaxies in the local Universe may 
approximate galaxies in an early stage of their formation. Later, 
Thuan \& Martin (1981)
identified blue compact dwarf (BCD) galaxies as a class of 
low-luminosity ($M_B$ $\ga$ --18) extragalactic
systems undergoing a strong burst of star formation (SF) while
exhibiting extremely low heavy element abundances.

For many years the best candidate for a truly young galaxy has been
I~Zw~18 $\equiv$ Mkn~116, with a metallicity $Z$ of about 1/50 $Z_{\odot}$
(e.g. Lequeux et al. 1979). Studies of this galaxy in the last decade
suggest indeed that it is chemically unevolved. Its
extremely low metallicity derived from the HII region-like spectrum
has been confirmed by many subsequent studies (see Izotov \& Thuan
(1998) and references therein).  Observations in neutral hydrogen (HI)
show that it is a very gas--rich system (Lequeux
\& Viallefond 1980; Viallefond et al.\ 1987; van Zee et al.\
1998a). Also, its HI gas appears to have an extremely low metallicity
(Kunth et al.\ 1994), although this last point is considered
controversial (Pettini \& Lipman 1995; van Zee et al. 1998a).

Until the end of the eighties, I~Zw~18 remained the only BCD with a
heavy element abundance as low as $Z_\odot$/50. Despite efforts by
many investigators to search for BCDs with equally low or lower
metallicities there remained a significant gap in metallicity between
I~Zw~18 and all other known BCDs.  This gap was partially filled when
Izotov et al.\ (1990), using the Russian 6--m telescope, tentatively
determined the BCD
SBS~0335--052 to have a metallicity lower than that of I Zw 18. However,
later Melnick et al. (1992) concluded that $Z$ = 1/41 $Z_{\odot}$ in SBS 
0335--052, making it
the second most metal--deficient galaxy known in the Universe.  This
BCD is now considered as one of the best candidates for a galaxy in
formation.

A dwarf galaxy experiencing its first burst of star formation 
should have an extremely low metallicity in its HII regions.
Such a galaxy should not show any sign of an older stellar population
produced in previous SF episodes. The gas mass would be expected to
significantly exceed that in stars.  Studies of SBS~0335--052 over the
last few years show that it appears to fulfill all those conditions
and that it is very probably a truly young galaxy. We briefly
summarize its main observational properties :

1. The oxygen abundance in the brightest HII region in the galaxy is
found to be 12 + log(O/H) = 7.30 $\pm$ 0.01 (with variations
from 7.15 to 7.33 on small scales), only slightly higher
than that in I~Zw~18 ( Melnick et al. 1992; Izotov et al. 1997;
Izotov et al. 1999).

2. Blue underlying, extended low--intensity emission is detected in
SBS~0335--052 on $V$, $R$ and $I$--band images.  The blue $(V-I)$ and
$(R-I)$ color distributions suggest that a significant contribution to
the emission of the extended low--intensity envelope is due to ionized
gas.  It is found that the observed equivalent width of H$\beta$
emission in the extended envelope is 2 -- 3 times weaker than the value
expected in the case of pure gaseous emission.  This could be
explained by the presence of underlying stellar emission from coeval
A stars (Izotov et al.\ 1997; Thuan et al.\ 1997).

3. From the optical colors of the low surface--brightness component together 
with evolutionary synthesis models,
Papaderos et al. (1998) conclude that the age of the underlying
stellar population is less than $\sim$ 100 Myr, with a total mass of $\sim$
10$^7$\,$M_\odot$, comparable with the total mass of young blue stars
(Izotov et al. 1997). Near-infrared (NIR) colors after correction for the 
gas contribution are consistent with a stellar population not older 
than 4 Myr and the possible contribution to the NIR light 
from an evolved stellar population in SBS 0335--052 cannot exceed
$\sim$ 15\% (Vanzi et al. 2000).  

4. No evidence for strong, narrow Ly$\alpha$ emission is found in a UV
spectrum obtained with the {\sl Hubble Space Telescope} ({\sl HST}) 
(Thuan \& Izotov 1997).  Instead, a damped
Ly$\alpha$ absorption line is observed which implies an HI column
density $N$(H\,I) = 7 $\times$ 10$^{21}$ cm$^{-2}$, the largest observed so
far for a BCD and comparable to the highest column densities observed
in Ly$\alpha$ clouds in the direction of quasars. The resonant O\,I
$\lambda$1302 line is also detected. Assuming that the O\,I line is
unsaturated and that it originates in the neutral gas leads to an
oxygen abundance in the neutral gas of $\sim$ 1/37000 of the solar
value, or about 1/900 of that in the HII regions around the young
super--star clusters. However, a more consistent interpretation of the
UV absorption lines of O, Si and S is that these lines originate not
in the HI but in the HII gas. If this interpretation is correct, then
the neutral gas in this system can be pristine and not polluted with heavy
elements at all (Thuan \& Izotov 1997).

5. Thuan et al. (1995) and Izotov \& Thuan (1999) have also argued on 
the basis of the very
small dispersion of the C/O and N/O ratios in extremely
metal--deficient BCDs with a metallicity less than 1/20 solar,
 that in those galaxies, C and N are made exclusively by
massive ($M$ $>$ 9 $M_\odot$) stars, because intermediate-mass (3
$M_\odot$ $\leq$ $M$ $\leq$ 9 $M_\odot$) stars have not had time to evolve and
release their nucleosynthesis products. Izotov \& Thuan (1999) suggest
that BCDs with $Z$ $<$ $Z_\odot$/20 are young, with ages less
than $\sim$ 100 Myr, the main--sequence lifetime of a 9 $M_\odot$
star being $\sim$ 40 Myr. Chemical abundances thus also suggest that
 SBS~0335--052, with $Z$ = $Z_\odot$/40, might be a young galaxy.

In view of the unusual properties of this candidate young galaxy we
considered it important to measure its neutral gas content and study
the HI structure and velocity field.  Measurements of the integrated
HI flux of this galaxy in the 21--cm line were carried out with the
Nan\c cay radio telescope (NRT) in 1993 (Thuan et al.\ 1999a). From that data
the total HI mass was estimated which indicates that this galaxy is
gas rich ($M_{\rm HI}$ $\sim$ 10$^{9}$ $M_{\odot}$) with less than a tenth
of its baryonic mass in stars.  The HI flux appeared to be high enough
to allow higher resolution mapping with the VLA. In this paper we
present the results of VLA observations of SBS~0335--052 and discuss
the properties of its HI density distribution and velocity field.

The paper is organized as follows. In section 2 we describe the
observations and data reduction. In section 3 we derive 
the HI density distribution and velocity field. In
section 4 we examine the gas and dynamical mass of the system and in
section 5 we focus on various aspects related to the star formation in
SBS~0335--052. The paper is summarized in Section 6.

%
%
\begin{deluxetable}{lcc}
\tablenum{1}
\tablecolumns{3}
\tablewidth{0pt}
\tablecaption{Observational parameters \label{tbl-1}}
\tablehead{
\colhead{Parameter} & \colhead{Value} & \colhead{Value}
}
\startdata
Date			& 26 Dec 94         	&   13 May 95		\nl
VLA configuration		&    C              	&      	\nl
No.\ of antennae	&    26             	&      26	\nl
Time on source [min]    &   163                 &      80        \nl
Central velocity [km\,s$^{-1}$]	&  4043  	&    4043	\nl
Half--power beam	& 20\farcs5$\times$15\arcsec\  &  61\arcsec$\times$45\arcsec\	\nl
Brightness temperature [K]& 2.0                   &   0.22       \nl
corresponding to 1 mJy\,beam$^{-1}$	& 	&		\nl \nl
Total bandwidth [MHz]	&    3.12 		&     3.12	\nl
Total no.\ channels	&   31 			&    31 	\nl
Channel width [km\,s$^{-1}$]	&   21.2 	&    21.2 	\nl
Noise in channel maps [mJy\,beam$^{-1}$]	& 0.5	&     1.0 \nl \nl
Total bandwidth [MHz]  	&    1.56		&     1.56	\nl
Total no.\ channels 	&   63			&    63		\nl
Channel width [km\,s$^{-1}$]    &    5.3 	&     5.3	\nl
Noise in channel maps [mJy\,beam$^{-1}$] & 1.0	&     2.0	\nl
\enddata
\end{deluxetable}

\section{Observations and Data Reduction}

The blue compact dwarf galaxy SBS~0335--052 was observed with the VLA in the
21--cm line of neutral hydrogen on December 26, 1994 for an 
on-source time of 163 min in the C
configuration, and on May 13, 1995 for an on-source time of 
80 min in the D configuration. The C
configuration, with baselines between 160 to 15000 wavelengths (or
0.03 to 3 km) was used 
to achieve high enough angular resolution in HI
to resolve features comparable in size to the optical size of the
galaxy ($\sim$15\arcsec). The D configuration, with baselines between 160
to 5000 wavelengths (or 0.03 to 1 km), was used to achieve high
surface brightness sensitivity for gas distributed at larger scales
(the 2$\sigma$ noise level corresponds to 7.5 $\times$ 10$^{19}$
cm$^{-2})$ and to search for companion HI clouds, such as those
detected near several other BCDs by Taylor et al. (1993, 1995, 1996a),
Taylor (1997). Table~\ref{tbl-1} summarizes the observations.

%
%
\begin{deluxetable}{lrr}
\tablenum{2}
\tablecolumns{3}
\tablewidth{200pt}
\tablecaption{Coordinates of reference stars used to tie the optical to 
the VLA coordinate frame \label{tbl-2}}
\tablehead{
\colhead{Star} & \colhead{$\alpha_{(1950.0)}$} & \colhead{$\delta_{(1950.0)}$}
}
\startdata
1  & 03$^{\rm h}$34$^{\rm m}$59\fs71 & --05\arcdeg13\arcmin04\farcs0   \\
2  & 03$^{\rm h}$35$^{\rm m}$02\fs64 & --05\arcdeg12\arcmin05\farcs9    \\
3  & 03$^{\rm h}$35$^{\rm m}$16\fs93 & --05\arcdeg13\arcmin19\farcs7   \\
4  & 03$^{\rm h}$35$^{\rm m}$19\fs23 & --05\arcdeg11\arcmin57\farcs7    \\
5  & 03$^{\rm h}$35$^{\rm m}$21\fs42 & --05\arcdeg10\arcmin29\farcs8   \\
6  & 03$^{\rm h}$35$^{\rm m}$26\fs86 & --05\arcdeg10\arcmin50\farcs6    \\
\enddata
\end{deluxetable}

The observations exploited the 4IF correlator mode. In this mode one
IF pair was used to take data over a bandwidth of 1.56 MHz, with 63
channels, giving a channel width of 24.4 kHz (5.3 km\,s$^{-1}$), while
the second IF pair was used over a 3.12 MHz bandwidth with 31
channels, each having a width of 97.7 kHz (21.2 kms$^{-1}$). The
higher velocity resolution provides detailed information on the
galaxy, whereas the lower velocity resolution data cover a larger
range in velocity space, at higher S/N ratio.

The data were edited and calibrated with the NRAO {\sc aips} package. The C
and D configuration observations were calibrated separately and
inspected. Once satisfied with the result, the data were combined and
Fourier transformed. In the end, data cubes were obtained measuring
512 $\times$ 512 pixels, each pixel measuring 5\arcsec\ on a side.  The
resulting synthesized beam was 20\farcs5 $\times$ 15\farcs0. The noise in 
the final cube, at a velocity resolution of 21.2 km s$^{-1}$,
is $\sim$ 0.5 mJy/beam in a single channel.

\section{HI Distribution and Velocity Field}

\begin{figure*}[tbh]
\figurenum{1}
\plotfiddle{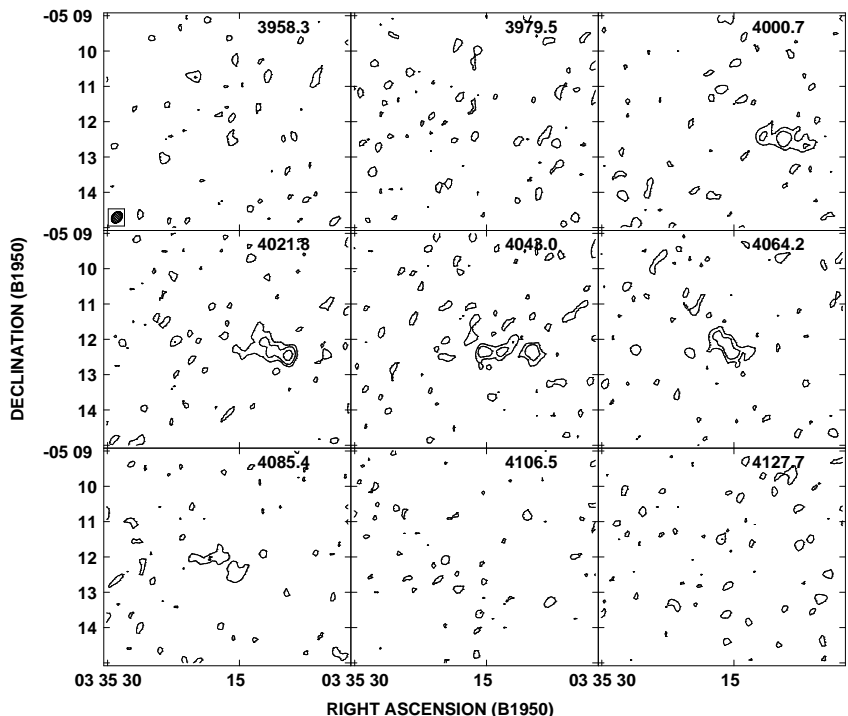}{0.cm}{270.}{50.}{50.}{-220.}{30.}
\vspace{9.cm}
\figcaption[Fig1.ps]{
Mosaic of channel maps of the low velocity resolution data showing
the HI distribution of SBS~0335--052.  Contours represent 1
(2$\sigma$), 2, 4 and 6\,mJy\,beam$^{-1}$.  The ellipse in the lower
left hand corner of the panel on the upper left indicates the beam
size.
\label{Fig1}}
\end{figure*}

\begin{figure*}[tbh]
\figurenum{2}
\plotfiddle{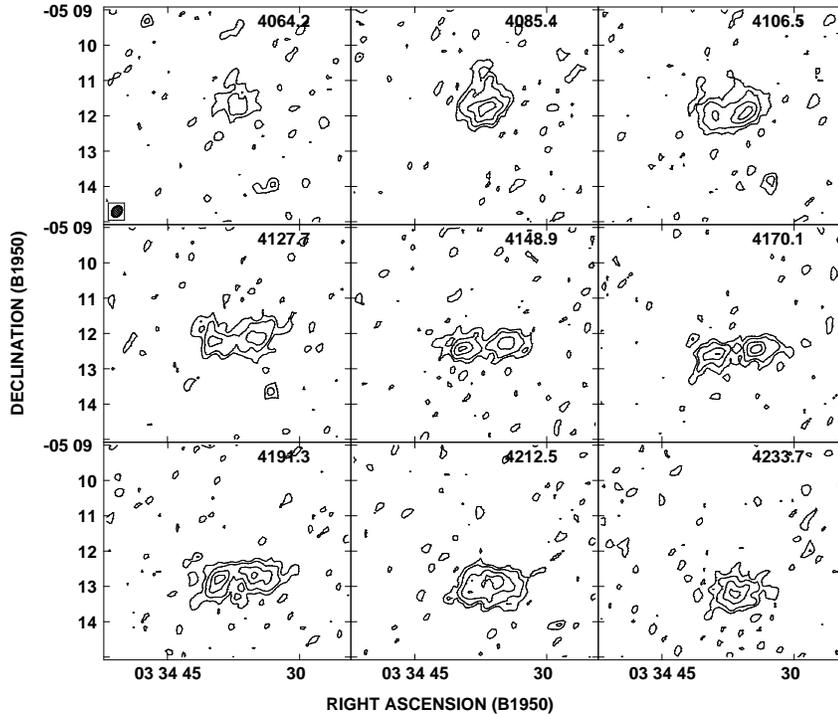}{0.cm}{270.}{50.}{50.}{-220.}{30.}
\vspace{9.cm}
\figcaption[Fig2.ps]{
Mosaic of channel maps of the low velocity resolution data showing
the HI distribution of the bright spiral galaxy NGC~1376.  Contours
represent 1 (2$\sigma$), 2, 4, 6 and 8\,mJy\,beam$^{-1}$.  The ellipse
in the lower left hand corner of the panel on the upper left indicates
the beam size.
\label{Fig2}}
\end{figure*}

In Figures~\ref{Fig1} and \ref{Fig2} we show the individual channel
maps over the velocity range where significant signal is detected. The
size of the synthesized beam is indicated by the ellipse in the lower
left hand corner of the first channel map in each of the two figures.
Fig.~\ref{Fig1} displays the HI emission
associated with SBS~0335--052 which is traced in the channels
corresponding to a range in velocity from 4000 to 4085 km\,s$^{-1}$.
Despite the low signal--to--noise in the channel maps, one can
appreciate the rather complex kinematics of the gas. 
There is an underlying rotation, with the 
eastern edge of the HI complex receding.  A prominent feature in the
data cube is a large, almost face--on spiral galaxy, NGC~1376. The
channel maps corresponding to the velocities covered by this object
are presented in Fig.~\ref{Fig2}. Its velocity ranges from 4064 to
4234 km\,s$^{-1}$. The central radial velocity, 4162\,km\,s$^{-1}$
(Rix \& Zaritzky 1995), is about 120\,km\,s$^{-1}$ higher than that
of our target galaxy.

Adopting a Hubble constant $H_{0}=$  75\,km\,s$^{-1}$ Mpc$^{-1}$, 
the redshift distance is 54.3 Mpc (Thuan et al. 1997),
so that 1\arcsec\ corresponds to
263\,pc. If NGC~1376 lies at the same distance as SBS~0335--052, their
projected distance of about 9\farcm5 then corresponds to 150
 kpc.

\begin{figure*}[tbh]
\figurenum{3}
\plotfiddle{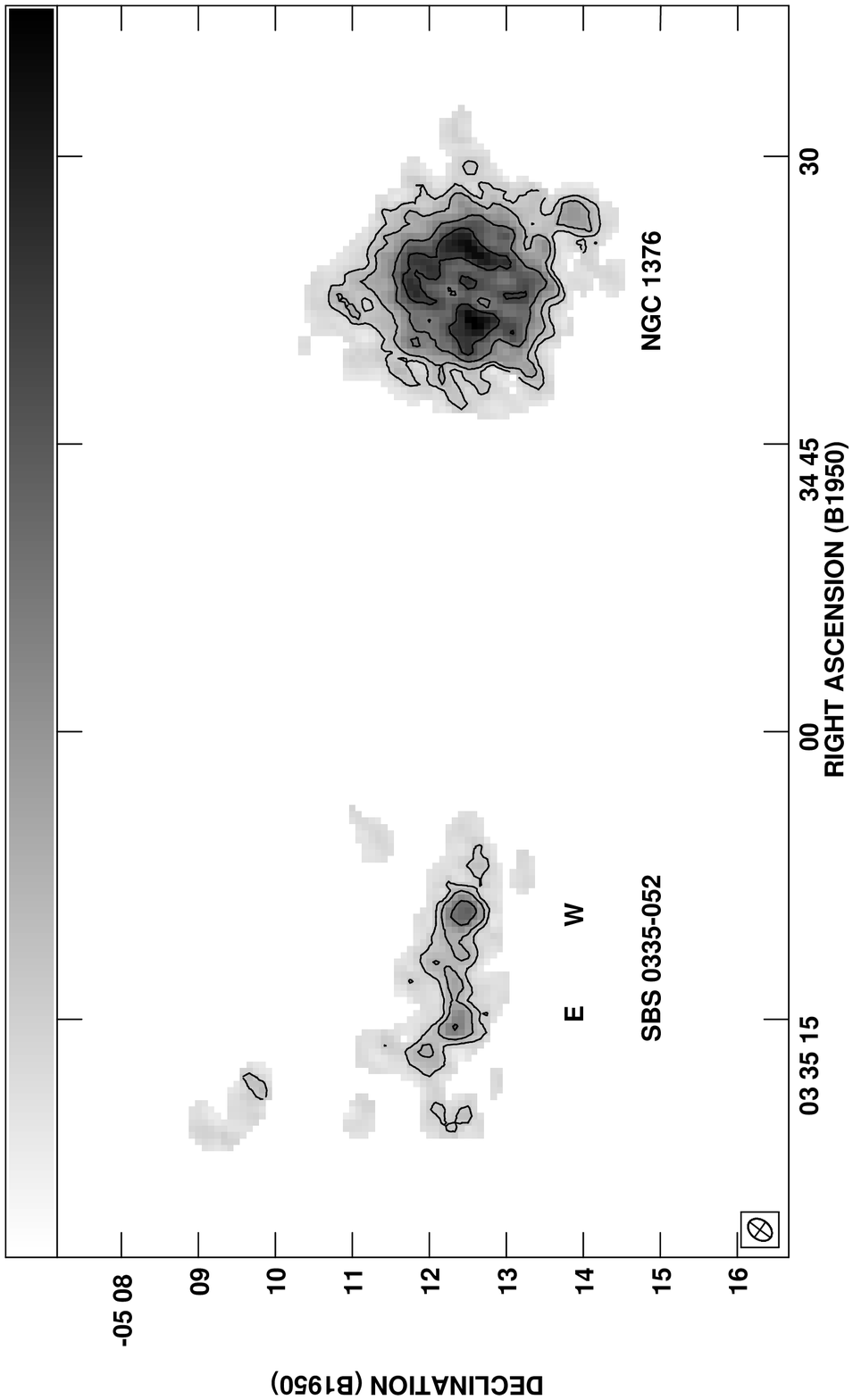}{0.cm}{270.}{50.}{50.}{-220.}{40.}
\vspace{8.cm}
\figcaption[Fig3_1.ps]{
Total HI map of SBS~0335--052 and companion galaxy NGC 1376,
based on the low velocity resolution data.  Contours represent 1.9
(5$\sigma$), 3.8, 7.6 and 11.4 $\times$ 10$^{20}$ cm$^{-2}$. The grey
scale is a linear representation of HI surface density ranging from 0 (white)
to 16 $\times$ 10$^{20}$ cm$^{-2}$. The ellipse in
the lower left hand corner indicates the beam size.
\label{Fig3}}
\end{figure*}

Fig.~\ref{Fig3} is a grey scale presentation of the integrated
intensity of the HI in the target field. It is obtained by creating a
zeroth moment map of a blanked data cube, i.e., a data cube where all
signal below the 2$\sigma$ level was deleted and in which only those
regions are preserved which show emission in at least two successive
channels. Both SBS~0335--052 and NGC~1376 are displayed. Notice the
elongated structure of the SBS~0335--052 HI cloud with two prominent
concentrations separated by about 84\arcsec\ (or
22\,kpc). These ``peaks'' are marked respectively by ``E'' and ``W'' on
the map. They are slightly resolved, as indicated by gaussian fitting
of the brightness distribution of small regions centered on these
peaks, and  are situated approximately symmetrically relative to the
cloud center.  The western peak is about a factor of 1.3 brighter than
its eastern counterpart.  The HI cloud as a whole is quite inhomogeneous.

%
%

%
%

 Both the E and W HI peaks are
embedded in a common, elongated envelope with a 
strong bend at the eastern edge.
The outermost contour in Fig. 3 
corresponds to a HI column density of $3 \times 10^{20}$
cm$^{-2}$ and a size of about 31 $\times$ 7 kpc.
At the 2$\sigma$ level above the noise (represented by the lightest grey scale 
tones in Fig. 3), which corresponds to a HI column 
density of 1.9 $\times$ 10$^{20}$ cm$^{-2}$,
the total extent of the HI emission is $\sim$
66 $\times$ 22 kpc.
Some additional weak HI features
may be present in the field around SBS~0335--052, including an
extended object to the NE. However, these detections are tentative at
best and should be confirmed with future higher S/N observations.  
The structure of the HI feature to the NE of the SBS~0335--052 HI 
complex at $\delta$ = --05$\arcdeg$ 09$\arcmin$ 30$\arcsec$ 
is quite unusual. No cases similar to this one were
found in studies devoted to a search for HI companions near HII
galaxies (Taylor et al.\ 1993, 1995) or low-surface-brightness (LSB)
 dwarf galaxies (Taylor et al.\ 1996b).

We should like to emphasize that an HI envelope of this size is
rare among BCDs and dwarf irregulars.  The typical size of HI
envelopes around BCDs is usually a few kpc (see for example the
discussion in Viallefond \& Thuan (1983) and in van Zee et
al. (1998b)). The only known HI envelope of comparable size is that of
HI~1225+01 (Chengalur et al. 1995).

\begin{figure*}[tbh]
\figurenum{4}
\plotfiddle{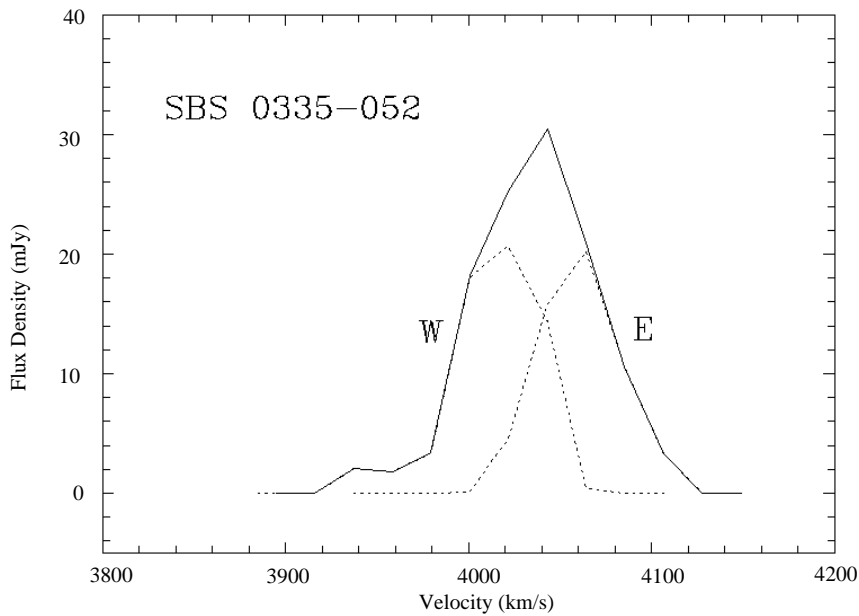}{0.cm}{270.}{50.}{50.}{-220.}{40.}
\vspace{8.cm}
\figcaption[Fig4_2.ps]{
Integrated profile of HI flux (in mJy) for the entire HI cloud (solid
line) in the system SBS~0335--052, as derived from the channel maps
displayed in Fig.\,1 for the low velocity resolution data.  The total
flux, integrated over this profile is 2.46\,Jy\,km\,s$^{-1}$. Dashed
profiles show integrated profiles for the E and W
components separately.
\label{Fig4}}
\end{figure*}

Figure~\ref{Fig4} represents the integrated profile of the HI cloud
associated with SBS~0335--052. It was calculated by integrating all the
flux in the region around SBS~0335--052, as seen in
Figure~\ref{Fig3}. The dashed lines are used to indicate separately
the integrated profiles of the eastern and western halves of the
HI cloud. Whereas the integrated profile width measured with the VLA
is close, within the errors, to that measured with the NRT
 (Thuan et al. 1999a), the VLA integrated flux of $2.46
\pm 0.18$
Jy\, km\,s$^{-1}$ is about a factor of two higher than that measured
with the NRT (1.28 $\pm$ 0.22 Jy\, km\,s$^{-1}$). This difference points
to a significant loss of integrated flux for the NRT due to its narrow
horizontal beam (FWHM = 3\farcm5) as compared to the larger
extent of up to $\sim$ 4\arcmin\ of the HI cloud in the EW direction.
After a first order
correction for the NRT beam shape, the NRT/VLA flux ratio 
 becomes $0.78 \pm 0.27$, which is consistent within the errors.

\begin{figure*}[tbh]
\figurenum{5}
\plotfiddle{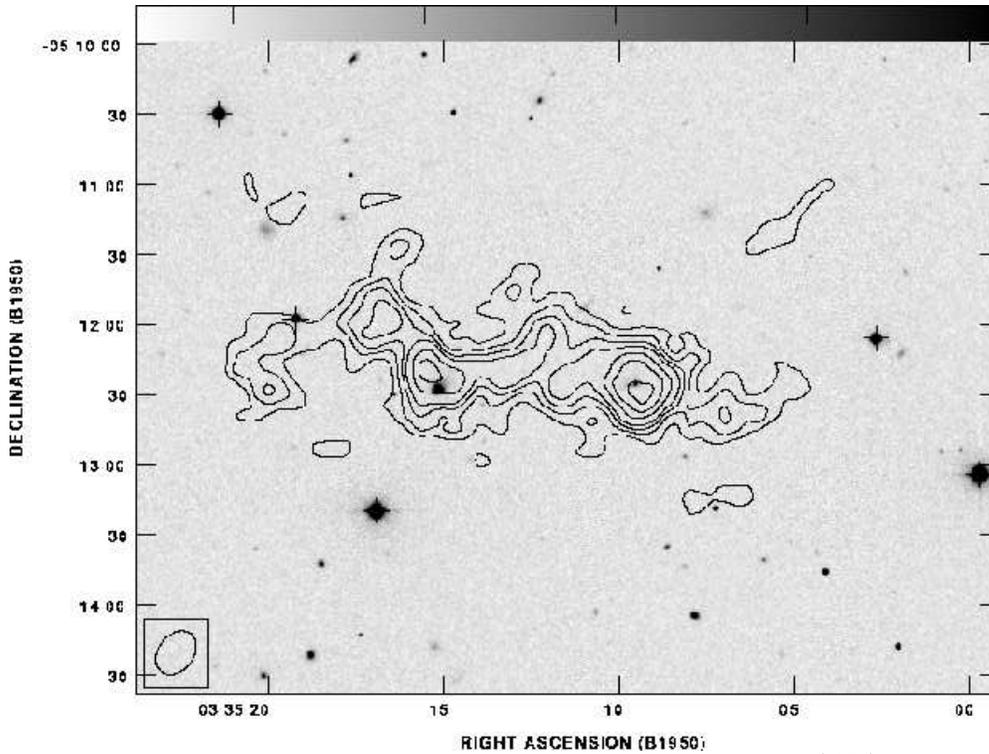}{0.cm}{270.}{60.}{60.}{-250.}{50.}
\vspace{9.5cm}
\figcaption[Fig5_1.ps]{
Overlay of the HI contours on top of an optical
$B$--band image from Papaderos et al. (1998).
The crosses mark the bright stars in the frame which have been used
to adjust the overlay. The brightness levels of the B image are presented 
on a linear scale and in arbitrary units. HI contours represent 0.75
(2$\sigma$), 1.8, 2.7, 3.6, 5.4, 7.2 and 9.0 $\times$ 10$^{20}$ cm$^{-2}$.
\label{Fig5}}
\end{figure*}

Fig.~\ref{Fig5} is an overlay of the HI contours on top of an optical
$B$--band image from Papaderos et al. (1998). The optical 
astrometry was done with the six stars marked by crosses. 
Their coordinates were obtained from the {\sl Hubble Space Telescope} Guide 
Star Catalog (Lasker et al. 1990) 
and are given in Table 2. As shown first by
Pustilnik et al.\ (1997) the two HI peaks are clearly identified with
two faint optical galaxies --- the eastern one with the object
SBS~0335--052 itself, and the western peak with a very compact, faint
dwarf galaxy with $m_B$ = 19.4.
A comparison of the optical and HI peak positions will be given in 
section 4.
 No 20 cm continuum emission was detected at the locations 
of SBS~0335--052, SBS~0335--052W or NGC 1376 at the 4 $\sigma$ 
level of 1 mJy. 
To avoid as much as possible
confusion, we will use the same nomenclature for the HI components as
for the optical counterparts, referring to the eastern HI component
simply as SBS~0335--052 and to the western one as SBS~0335--052W. When
referring to the entire HI complex we will employ the term SBS~0335--052
{\it system}.

\begin{figure*}[tbh]
\figurenum{6}
\plotfiddle{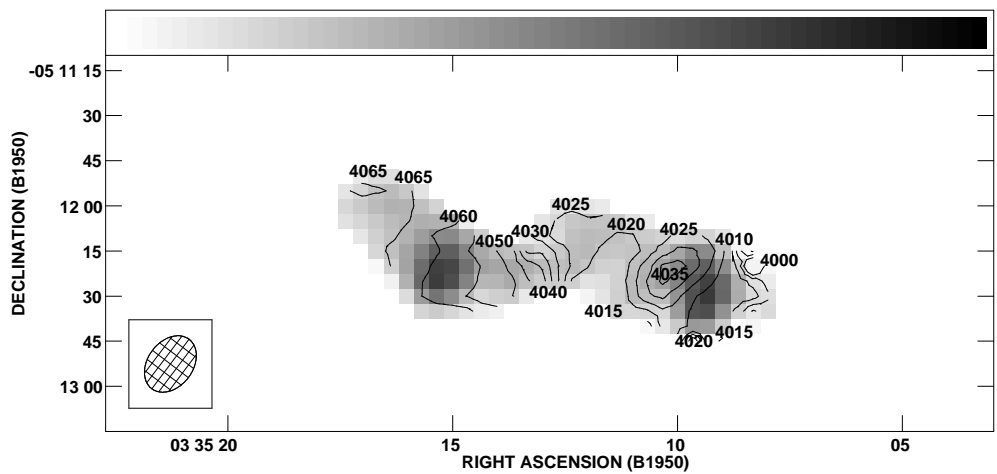}{0.cm}{270.}{50.}{50.}{-220.}{50.}
\vspace{7.cm}
\figcaption[Fig6_1.ps]{
Isovelocity lines of HI 
superimposed on the HI surface brightness map. The grey scale is a
linear representation of HI surface density ranging from 0 (white) to 
11 $\times$ 10$^{20}$ cm$^{-2}$.
\label{Fig6}}
\end{figure*}
\begin{figure*}[tbh]
\figurenum{7}
\plotfiddle{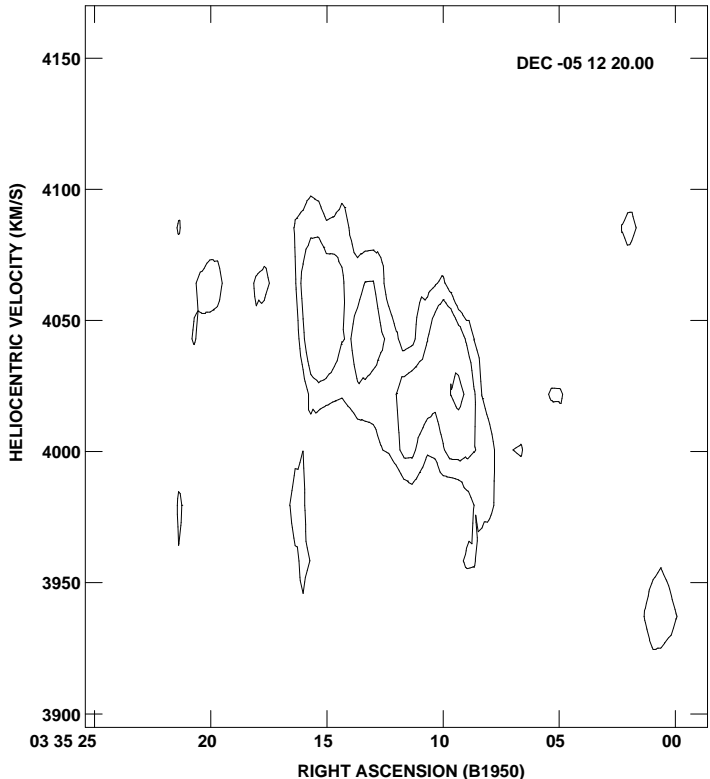}{0.cm}{0.}{50.}{50.}{-170.}{-335.}
\vspace{10.cm}
\figcaption[Fig7_2.ps]{
Position--velocity slice through the HI emission associated with 
SBS\,0335--052 along an east--west line, cutting through both HI 
emission peaks referred to as E and W in the text. Contours represent 
1.0 ($2 \sigma$), 2, 4, and 6 mJy\,beam$^{-1}$. The declination
along which the slice was taken is given in the top right corner.
\label{Fig7}}
\end{figure*}

In Fig.~\ref{Fig6} we show an enlarged image of the integrated HI flux
distribution in the SBS~0335--052 system, the grey scale being a linear
representation of the integrated HI flux, with superimposed
isovelocity lines with a contour interval of 5 km\,s$^{-1}$.
 The velocity field of the HI gas is quite complex. There
is a general gradient across the system from 4085 km\,s$^{-1}$ at the
eastern edge to 4000 km\,s$^{-1}$ on the western rim. However, there
are clear deviations from this general trend near the E and W
peaks. This is perhaps better illustrated in Fig.~\ref{Fig7} which is a
position--velocity (P--V) diagram made along a line through the two
peaks and based on the low velocity resolution data.
 If the HI complex is treated as a
single system, then there is a velocity gradient across, with a
maximum velocity difference of $\sim$ 80 km\,s$^{-1}$ between the two sides.
If it is considered however as two distinct HI clouds, then  
the velocity gradients are solid-body in nature, with
maximum velocity differences of $\sim$ 35--40 km\,s$^{-1}$ within each
cloud.


%
%

%
%

%
%

%
%

%
%


%
%
\begin{deluxetable}{lccc}[hbtp]
\tablenum{3}
\tablecolumns{4}
\tablewidth{450pt}
\tablecaption{Some observed and derived characteristics of the
SBS~0335--052 system \label{tbl-3}}
\tablehead{
\colhead{Parameter} & \colhead{0335--052W} & \colhead{0335--052E}
& \colhead{References}
}
\startdata
$\alpha_{(1950.0)}$ (opt)\tablenotemark{a}
		& 03$^{\rm h}$35$^{\rm m}$09\fs53
&  03$^{\rm h}$35$^{\rm m}$15\fs17  &         4          \\
$\delta_{(1950.0)}$ (opt)			
		&--05\arcdeg12\arcmin25\farcs2
& --05\arcdeg12\arcmin27\farcs6     &         4           \\
$B$ [mag]   &   19.37          &   17.00      &   1    \\
$L_B$ [10$^{8}$ $L_B{\odot}$]  &  0.82  &  7.3      &   1 \\
optical angular size [\arcsec]\tablenotemark{b}    &  14$\times$14 & 23$\times$20       &    1       \\
optical linear size [kpc]                          & 3.7$\times$3.7& 6$\times$5.3  & 1 \\
$V_{opt}$ [km\,s$^{-1}$]   
		& 4069$\pm$20  & 4060$\pm$12   &   2,3     \\
$V_{\rm HI}$  [km\,s$^{-1}$]   & 4017$\pm$\ 5  & 4057$\pm$\ 5   &   4    \\
$\alpha_{(1950.0)}$ (HI)
		& 03$^{\rm h}$35$^{\rm m}$09\fs50 	
& 03$^{\rm h}$35$^{\rm m}$15\fs21      &                  4       \\
$\delta_{(1950.0)}$ (HI)
		& --05\arcdeg12\arcmin25\farcs5 	
& --05\arcdeg12\arcmin21\farcs5        &                  4       \\
Peak $N_{\rm HI}$ [$10^{20}\,{\rm cm}^{-2}$] & 9.98  &    7.42   &    4        \\
Peak HI mass surface density [$M_\odot\,{\rm pc^{-2}}$]  
					&  8.0  &    6.0    &     4       \\
$M_B^0$                                 & --14.3 &  --16.7    &     1       \\
HI angular size [\arcsec]\tablenotemark{c} &  128$\times$67 &  122$\times$83  &     4        \\
HI linear size [kpc]                    &  33.7$\times$17.6 & 32.1$\times$21.8  & 4 \\
$M_{\rm HI}$ [$M_{\odot}$/10$^{8}$]  & 8.9 &  8.0     &     4        \\
$M_{\rm HI}$/$L_B$ [$M_{\odot}$/$L_B{\odot}$]  & 10.8 &  1.1 &  4        \\
$M_{tot}$/$L_B$ [$M_{\odot}$/$L_B{\odot}$] & 26.8 & 2.2 &     4        \\
\enddata
\tablenotetext{a}{Optical coordinates derived from DSS-2 images.}
\tablenotetext{b}{Optical size corresponding to the isophotal level 
of 27.5 mag arcsec$^{-2}$ in $B$ (Papaderos et al.\ 1998).}
\tablenotetext{c}{Maximum
 observed  extent of the associated HI gas.}
\tablenotetext{d} {$L_{B\odot}$ is the solar blue luminosity corresponding
  to M$_{B\odot}$=5.48}
\tablerefs{1. Papaderos et al.\ (1998). 2. Lipovetsky et al. \ (1999).
3. Izotov et al.\ (1997). 4. This paper.}
\end{deluxetable}

\section{Results}


As mentioned in the previous section, we derive an
integrated HI flux over the entire region around the cloud of
2.46\,Jy\,km\,s$^{-1}$. This corresponds to a
total HI mass: $M_{\rm HI}$ = 1.68 $\times$ 10$^{9}$
$M_{\odot}$. Since the blue luminosities of the 
eastern and western galaxies are respectively
7.3 $\times$ 10$^{8} L_{\odot}$ and 0.82 $\times$10$^{8} L_{\odot}$, 
the $M_{\rm HI}$/$L_B$ of the SBS~0335--052 system is of order 2.1. 

Accounting for a mass fraction of 0.245 for helium
(Izotov \& Thuan 1998), the total gas mass in the system is:

$M_{gas} =$  2.1 $\times 10^{9}$\,$M_{\odot}$.

 With a Salpeter initial mass function, and lower and 
upper mass limits of 0.8 and 120 $M_{\odot}$ respectively and using the 
models of Schaerer \& Vacca (1998), Papaderos et al. (1998) have 
derived a total stellar mass for the starburst and underlying stellar 
components of 3.1$\times$10$^6 M_{\odot}$. Lowering the lower mass limit 
to 0.1 M$_{\odot}$ would increase the total stellar mass to 
$\approx$3.7$\times$10$^{7} M_{\odot}$. From mid-infrared observations, 
Thuan et al. (1999b) have found that as much three-quarters of the current 
star formation activity in SBS 0335--052 can be hidden by dust. Correcting for 
this effect would bring the total stellar mass to 7.3 $\times$ 10$^{6} 
M_{\odot}$ and 8.8 $\times$ 10$^{7} M_{\odot}$ for lower mass cut-offs of 
0.8 and 0.1 $M_{\odot}$ respectively.   
     

Since SBS~0335--052W is about one order of
magnitude less luminous and slightly less chemically evolved,
its stellar mass is expected to be about 10 times lower than that of the 
eastern
galaxy. Thus the total stellar mass in both dwarf galaxies is at most
10$^8 M_{\odot}$, or no more than 5\% of the total gas mass.


The two peaks in the distribution of the integrated HI map are partly
resolved with the synthesized beam of 20\farcs5 $\times$ 15\farcs0 (or $ 5.4
\times 3.9$\,kpc at the adopted distance of 54.3\,Mpc). The position
angle of the beam is PA = --37\arcdeg. We used the {\sc aips} task
{\sc imfit} to estimate the properties of both HI peaks.  The central
brightness of the western peak is 25.8\,K\,km\,s$^{-1}$,
and its FWHM size after deconvolution with the beam is (26\farcs7 $\pm$
1\farcs5) $\times$
(15\farcs6 $\pm$ 1\farcs5) at PA = 53\arcdeg\ $\pm$ 7\arcdeg.  For the
eastern peak we get a central brightness of
19.2\,K\,km\,s$^{-1}$, and a FWHM size after
deconvolution of (28\farcs4 $\pm$ 1\farcs7) $\times$ (12\farcs7 $\pm$
4\farcs9) at PA = 53\arcdeg\ $\pm$ 9\arcdeg.  These central brightnesses
correspond to a column density of HI:

$N_{\rm HI}$(west) = 10.0 $\times$ 10$^{20}$ cm$^{-2}$ = 8.0 $M_{\odot}$ pc$^{-2}$,

$N_{\rm HI}$(east) = 7.4 $\times$ 10$^{20}$ cm$^{-2}$ = 6.0 $M_{\odot}$ pc$^{-2}$.

The faintest regions of the outer HI disk (roughly corresponding to a
2$\sigma$ noise level over two consecutive channels) correspond to
column densities of 7.5 $\times$ 10$^{19}$ cm$^{-2}$ or about 0.6
$M_{\odot}$ pc$^{-2}$.  Both HI concentrations are elongated, oriented
at the same position angle, roughly SW to NE. Their FWHM angular sizes
are very similar and correspond to a linear extent of about
7.1 $\times$ 3.7 kpc. In the SW--NE direction the intrinsic
(deconvolved) width of both HI peaks is about a factor of two larger
than that of the VLA beam, but in the perpendicular direction it is
only 0.7 -- 0.8 of the beam width.

The masses of the eastern and western components, simply dividing the
entire cloud in two halves, as was done in Figure~\ref{Fig4}, leads to
gas masses (corrected for He) of:

$M_{gas}$(west) = 1.11 $\times 10^{9}$\,$M_{\odot}$,

$M_{gas}$(east) = 0.99 $\times 10^{9}$\,$M_{\odot}$.

The peak column density, measured for the E--peak, $7.4 \times
10^{20}$\,cm$^{-2}$ is a factor of 10 lower than the value derived
from Ly$\alpha$ absorption based on {\sl HST} observations (Thuan \& Izotov
1997). This is also a factor 4 less than the peak brightness of HI for
I~Zw~18 as derived from VLA observations by van Zee et al. (1998a)
with a synthesized beam $\sim$ 5\arcsec\ (corresponding to a linear
resolution of 0.25 kpc). Both facts seem to indicate that despite the
fact that some fraction of the brightest parts of the SBS~0335--052 HI
complex is resolved by our VLA beam, structure at
scales significantly smaller than the beam must exist and that the
structure of the neutral gas in the vicinity of the region with high
current SF rate is clumpy, explaining the difference in
the peak HI column density derived from fitting the damped Ly$\alpha$ line 
profile (Thuan \& Izotov 1997) and that derived here. Probably, an
inhomogeneous gas distribution due to HII regions and SNR shells
affect the structure of the neutral ISM. Such shell structure is
already hinted at by the optical image obtained by Melnick et
al.\ (1992) and confirmed on {\sl HST} images by Thuan et al. (1997).

The coordinates of the HI peaks, determined with the {\sc aips} task
{\sc imfit}, as defined by the centers of gaussian components, are
presented in Table~\ref{tbl-3}. Their r.m.s. uncertainties are
0\fs18 $\times$ 1\farcs3 for the eastern HI peak, and
0\fs11 $\times$ 0\farcs8 for the western one. 

The optical position of SBS~0335--052W coincides within the errors with 
the position of the corresponding HI peak.
For SBS~0335--052, while there is no shift
 in right ascension between the optical and HI peaks 
within the cited errors, there exists a difference in declination
 $\Delta\delta$ =
+6\farcs1. This is more than twice the combined error in declination 
and seems significant. To check the reality of this offset,
we have also compared the optical image with a map made only with the 
C-configuration data. The displacement is still present.  
Taken at face value, the shift between the optical and HI peaks
for SBS~0335--052 is
about 1.6 kpc, which
corresponds to the offsets reported by, e.g., Viallefond \& Thuan (1983). 
Note, 
however that van Zee et al. (1998a,b), with higher spatial resolution HI maps,
find no significant offset between the peaks of the HI and optical intensity 
distributions in the BCD I~Zw~18. 


Spectroscopy of the newly discovered galaxy at the position of the
western peak with the 6--m (Pustilnik et al. 1997),
and with the Multiple Mirror and Keck II telescopes 
(Lipovetsky et al. 1999) reveals an
emission--line spectrum with a radial velocity close to that of the HI
cloud. SBS~0335--052W, with an absolute magnitude $M_B$ about $-$14.3
is,
like SBS~0335--052, an extremely metal--deficient object. Its
metallicity is 1/50 that of the sun, comparable to that of I Zw 18.


The total dynamical mass is difficult to determine as this depends
crucially on one's interpretation of the SBS~0335--052 system. In this
paper we will present two strawman hypotheses. The first one is based
on the assumption that the system is in fact one cloud with two
condensations. The second hypothesis is that we are witnessing the
interaction of two systems and that the observed HI shows a tidal
bridge between both optical galaxies and tidal tails stretching
towards the northeast and to the west. 

In the case of one giant cloud, we can estimate the total
gravitational mass by equating the centrifugal and gravitational
forces at the edges of the disk. We take for the maximum rotational
velocity 40 km\,s$^{-1}$, and a radius of the disk equal to 16 kpc (Fig. 7).
The resulting estimated lower limit for the total dynamical mass is:

$M_{dyn}$($R$ $<$ 16 kpc) = 5.9 $\times$ 10$^{9}$ $M_{\odot}$. 

In the case of two interacting clouds we estimate, based on the
position--velocity diagram (Fig. 7) a full width to zero
velocity range of 65\,km\,s$^{-1}$ for SBS~0335--052 and
75\,km\,s$^{-1}$ for SBS~0335--052W. The linear sizes of the HI clouds
are at most 12\,kpc (this is using the full extent rather than the
FWHM along the major axis of each object). We then find:

$M_{dyn}$(east) = 1.6 $\times$ 10$^{9}$ $M_{\odot}$, 

$M_{dyn}$(west) = 2.2 $\times$ 10$^{9}$ $M_{\odot}$, 

\noindent or a total of $M_{dyn}$ (east+west) = 3.8 $\times$ 10$^{9}$ $M_{\odot}$,
64\% of that derived on the basis of the more extreme hypothesis
of the objects forming one cloud. It should be noted that these
estimates are strict {\it lower} limits as the inclination is
completely unknown and was assumed to be 90\arcdeg\ (or edge--on).

 We have shown that in both hypotheses, 
we need to invoke substantial amounts of dark matter in order to
reconcile the observed baryonic content with the dynamical masses
derived on the basis of the observed kinematics.
To obtain yet another (lower) limit to the amount of dark matter which
might be hiding within the SBS~0335--052 system, we can take the observed
(projected) separation and radial velocity difference and calculate
the mass which this implies assuming the two clouds to be in a bound
orbit. This works out to be $M_{tot} = 9 \times 10^9$\,$M_\odot$ (see
van Moorsel 1987 for details). Therefore, both the individual clouds,
as well as the SBS~0335--052 system require substantial amounts of
non--visible matter to make up for the inferred masses as based on
their dynamics. 

\section{Discussion}

\subsection{SBS~0335--052 in relation to other candidate young galaxies}


It is interesting to compare this system with other candidate young
galaxies.
 There are three galaxies in the literature with similar
characteristics which have been proposed as possible young
objects: I~Zw~18, ESO~400$-$G43 and HI~1225+01.

The HI in I~Zw~18 has a complex structure with
three distinct HI components (Viallefond et al.\ 1987; van Zee et al.\
1998a), two of them associated with star--forming regions: the NW and
SE components in the main body of the galaxy and component C to the
north (e.g. Dufour et al.\ 1996).  But unlike SBS~0335--052, where the
two centers of star formation are well separated spatially, the three
star-forming regions in I Zw 18 are connected causally, star formation
self-propagating from component C to the SE component (Izotov \& Thuan
1998).

In the case of ESO~400$-$G43 (Bergvall \& J\"ors\"ater 1988), the two
HI components are well separated by a projected distance of $\sim$ 40
kpc, each component containing an optical counterpart. Only one of the
optical galaxies has been studied in detail.  It has been described as
a possible young galaxy, although its metallicity $Z$ = $Z_{\odot}$/8,
is probably
too high for it to be forming its first generation of stars (see
Izotov \& Thuan 1999).

The HI~1225+01 system (Salzer et al.\ 1991; Chengalur et al.\ 1995),
at a distance of 20 Mpc, presents several similarities
to the HI envelope associated with SBS~0335--052. It also
has two extended components, separated by a projected distance of
98 kpc, and with an HI tidal bridge linking them. A dwarf HII
galaxy 12 kpc in size is located within the 40 kpc NE HI
component, approximately at the position
of the HI peak. The total gas mass of the NE HI component of $\sim
3 \times 10^{9}$ $M_{\odot}$, is close to that of the HI gas in the
SBS~0335--052 system. The dynamical mass is a factor of 3 larger
than the gas mass. Due to a smaller inclination angle ($i \sim$
30\arcdeg) and better spatial and velocity resolution the authors
are able to distinguish bar--like and spiral structure in the NE HI
component. 

The very low metallicity, $Z$ $\sim$ 1/20 $Z_{\odot}$, in the HII region
associated with the current SF burst and the very low mass fraction in the 
form of stars suggest that the HI~1225+01 system probably is in the process of
forming its first generation of stars.  One of the arguments by Salzer
et al.\ (1991), who claim that the current SF burst in this HII galaxy
is possibly not the first one, is that there is a significant
abundance of nitrogen, which the authors assume to be produced by
intermediate--mass stars.  However, as mentioned earlier, the finding
of a constant N/O ratio 
with a very small dispersion in the most metal--poor BCDs, those
with $Z$ $<$ $Z_\odot$/20 (Thuan et al. 1995; Izotov \& Thuan 1999)
strongly suggests primary production of nitrogen by massive stars in
galaxies as metal-deficient as SBS~0335--052, and thus there is no
need for a second burst of star formation in the optical counterpart
of HI~1225+01 to account for the nitrogen abundance.  The second HI
component has no detectable optical emission, but shows a velocity
field corresponding to that of an inclined rotating disk.  Thus,
Chengalur et al. (1995) suggest that the most probable interpretation
for this system is a SF burst in a primordial HI cloud (the NE
component) due to the tidal action of another massive HI cloud.



The HI properties of the SBS~0335--052 system also show some similarities with
those of II~Zw~40 (Brinks \& Klein 1988; van Zee et al.\ 1998b).
 The metallicity of this BCD is $\sim Z_{\odot}$/6, 
considerably higher than those of the above-mentioned systems, and 
suggesting that this galaxy is more 
evolved than SBS~0335--052 and not necessarily its analog. Nevertheless
it is instructive to briefly compare the two objects to discuss their 
different evolutionary states.
The II~Zw~40 system is most likely a merger, showing two HI tidal tails of 
$\sim$ 6.5 and 15 kpc in size. The extent of the HI gas and the HI 
masses of the tails are reminiscent of the situation in SBS~0335--052.
However, the optical morphologies of SBS~0335--052 and II~Zw~40 differ.
Whereas II~Zw~40 resembles an advanced merger with optical 
tidal tails (van Zee et al. 1998b),   
SBS~0335--052 exhibits two regular-looking still 
widely separated dwarf galaxies which are perhaps in an earlier stage 
of what could evolve into a merger. Over time, SBS~0335--052 might very well 
develop into a system resembling II~Zw~40.

\subsection{Possible star formation triggers}

One of the unsolved problems ever since the discovery of the first BCDs
(Sargent \& Searle 1970) concerns
the triggering mechanism for the star formation. And perhaps even more
baffling, why has star formation not occurred earlier in the lifetime
of these objects. 

In all likelihood, tidal triggering is probably responsible for the 
current star formation in the SBS 0335--052 system. 
We shall again consider two hypotheses: 1) the case where 
SBS 0335--052 is one huge HI cloud containing two 
star-forming centers; and 2) the BCDs are the nuclei of two distinct 
interacting HI clouds.  

In the case of a single self-gravitating HI cloud, the tidal triggering 
is probably due to the companion massive galaxy NGC~1376 whose 
HI map is seen in Figure 8. The resolution of the map is too low 
to say much about the HI morphology. However we can note that the HI 
gas is, as usual, more extended 
than the optical disk, and that there is a conspicuous 
lack of HI in the nuclear region of the galaxy.  
A lower limit to the distance between the galaxy and the SBS~0335--052 system
  is their observed projected distance of 150 kpc.
NGC~1376 is an Scd galaxy with $m_{pg} = 12.8$, $M_{B} = -21.0$, an 
optical diameter $D$ = 2\farcm0 and an inclination of 
21\arcdeg. Its HI profile is characterized by a full width at 20\%
of peak intensity of 179 \,km\,s$^{-1}$ and a heliocentric
 radial velocity $V_{\rm HI}$ = 4162 km\,s$^{-1}$.
It is a member of a medium density group listed as
LGG 103 (Garcia 1993) and composed of at least 14
galaxies. NGC~1376 is situated on its south--western outskirts.
The closest known group member is at a projected distance of about 0.7
Mpc.
Rix \& Zaritzky (1995) note that NGC 1376 has 3 spiral arms, and Elmegreen
\& Elmegreen (1987), in their classification of spiral galaxies,
assign this galaxy to class 2, which means that its arms are
``fragmented spiral pieces with no regular pattern''. This suggests that the 
spiral has recently experienced a tidal disturbance. 

The VLA HI data does point to a small galaxy, companion to NGC 1376,
that may be at the origin of that tidal disturbance. 
Fig. 8 shows a small irregular HI blob SW of the main body
of NGC~1376 at $\alpha =  03^{\rm h} 34^{\rm m} 32.9^{\rm s}$ and 
$\delta$ = --05$\arcdeg$13$\arcmin$51$\arcsec$. Its detection is weak, but 
fairly reliable as it can be seen in at least two neighboring channel maps,
at a heliocentric velocity of  4117$\pm$5 km\,s$^{-1}$. Its integrated flux 
density is 0.17 Jy km s$^{-1}$, which corresponds to an HI mass of 
1.2 $\times$ 10$^{8}$ $M_\odot$ at the distance of 54.3 Mpc.   
This blob is kinematically
detached from the regular velocity field of the galaxy and can be identified 
optically on the Deep Sky Survey (DSS) with what appears to be a faint 
elongated dwarf companion, not listed in NED and with a diameter of  
$\sim$ 23\arcsec (i.e. a linear size of $\sim$ 6 kpc) 
at a  projected distance of 27.5 kpc from NGC 1376. The HI profile 
of the dwarf galaxy is two 
channels wide so that its velocity width is $\sim$ 42 km s$^{-1}$. 
This gives a lower limit 
(the inclination being unknown) 
of $\sim$ 2.5 $\times$ 10$^8$ $M_\odot$ for its dynamical mass.

%
%

\begin{figure*}[tbh]
\figurenum{8}
\plotfiddle{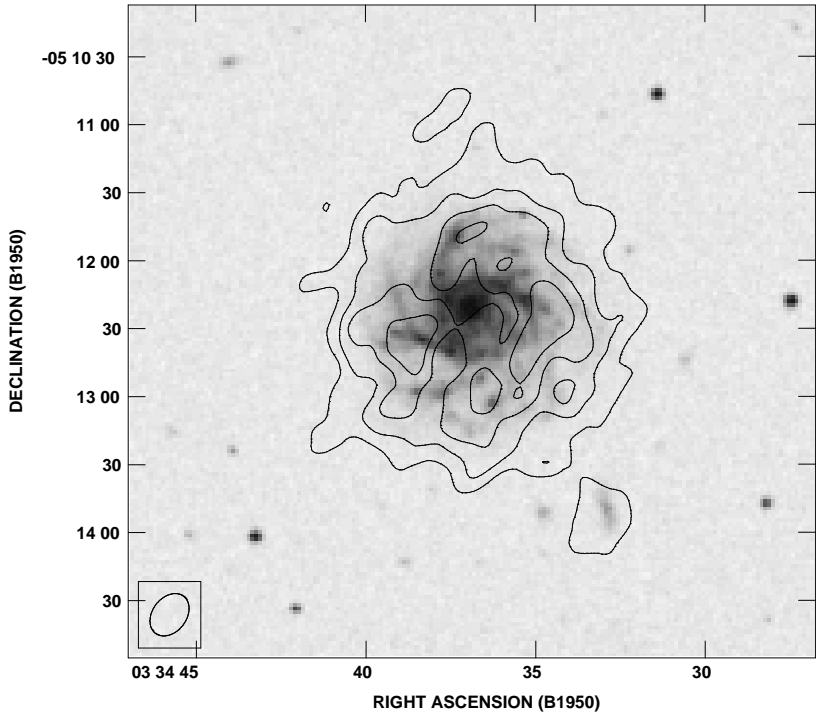}{0.cm}{270.}{50.}{50.}{-220.}{30.}
\vspace{9.cm}
\figcaption[Fig9_2.ps]{
Overlay of the HI contours on top of a DSS image of NGC 1376. Contour levels 
are at 3.8, 7.6, 11.4  and 15.2 $\times 10^{20}$ cm$^{-2}$.
There is a small HI blob SW of NGC 1376 which coincides with a 
low-surface-brightness dwarf galaxy not listed in NED.
\label{Fig8}}
\end{figure*}


The total dynamical mass of NGC~1376, as estimated from the width of
the HI profile (inclination corrected to $V_{rot}$ = 250
\,km\,s$^{-1}$) and the radius of the HI distribution, is about $5
\times 10^{11}$\,$M_{\odot}$. Taking into account the results by
Zaritsky \& White (1994) and Zaritsky et al.\ (1997) on distant dwarf
satellites of spiral galaxies and masses of their dark matter halos, the total
gravitational mass of NGC~1376, exerting a tidal force on
SBS~0335--052, can be up to twice that value, or $10^{12}$\,$M_{\odot}$.
A detailed understanding of the triggering mechanism of a SF burst due to
tidal forcing such as due to NGC~1376 is still lacking. It will
require a more careful study and modelling which is beyond the scope
of this paper. However, it is clear from some simple estimates, that
this massive spiral galaxy is close enough to the HI complex to induce
in it a gravitational instability via mechanisms such as those
suggested by Icke (1985), Noguchi (1988) and Olson \& Kwan (1990a,b). 
The mechanism proposed by Noguchi (1988) generates a central bar,
which in turn causes gas to sink towards the center of the disk and
induce its collapse.  No observational evidence is seen, neither for a
central bar--like structure in the HI cloud of SBS~0335--052, nor for
a central density concentration. The mechanism by Olson \& Kwan (1990a,b)
works via a large tidal increase of the inelastic collision rate of
individual gas clouds, and their merging and collapse. It is basically
a stochastic process. 

The most probable mechanism is that proposed by Icke (1985) involving 
tidal acceleration to supersonic velocities of gas layers in the external 
parts of an HI disk, with subsequent
dissipation of the kinetic energy and loss of dynamical stability of
the gas disk.
 It takes only a moderate tidal force, and acts from
larger distances in comparison with the two other mechanisms.
Adopting a radius $R$ = 16 kpc for the HI disk and a circular 
velocity $V$ of 40 km\,s$^{-1}$ at $R$, a dynamical mass within $R$
of 6 $\times$ 10$^9$ $M_\odot$ for the  SBS~0335--052 system, a total 
dynamical mass of 10$^{12}$ $M_\odot$ for NGC 1376, and a sound 
speed of the HI gas of 10 km\,s$^{-1}$, we find that for pericenter distances
smaller than $\sim$ 412 kpc,
the gas in SBS~0335--052 can be accelerated to supersonic velocities.
This distance is
significantly larger than the present projected distance of 150 kpc 
 so that
sufficiently strong tidal forces of NGC~1376 on the HI complex are expected.
Such a tidal triggering 
mechanism may explain  the fact that the two star-forming centers 
began nearly simultaneously, and are aligned along the line joining the 
centers of the two objects, symmetrically on either 
side of the center of the SBS~0335--052 system, just as the tidal forces of 
the moon on the earth causes high ocean tides on diametrically opposite 
locations on earth.

Alternatively, we may consider that we do not have a single huge HI complex,
but two distinct smaller HI components interacting tidally with each other 
and triggering star formation at the location of the two HI peaks.
In this case, the tidal force $F_1$ exerted by each cloud is proportional to 
$M$ (cloud) / $d_1^3$ where  $d_1$ is the projected 
separation between the two BCDs. On the other hand, the tidal force $F_2$ 
exerted by 
NGC 1376 is proportional to $M$ (NGC 1376) /  $d_2^3$
where  $d_2$ is the projected separation between the galaxy and 
the SBS~0335--052 system. Adopting $M$ (cloud) = 2 $\times$ 10$^9$ $M_\odot$,
$M$ (NGC 1376) = 10$^{12}$ $M_\odot$,  $d_1$ = 22 kpc and  $d_2$ = 
150 kpc, we find $F_1$ / $F_2$ = 0.6, i.e the tidal forces in the two 
cases are comparable and we cannot say for sure which of the two scenarios 
discussed above prevails. In the case of two distinct HI components, we 
find that the Icke (1985) mechanism will apply for pericenter distances 
less than $\sim$ 27 kpc.
 
Clearly, detailed hydrodynamical calculations of the tidal interaction 
between two
HI clouds and between a massive galaxy like NGC~1376 and a primordial gas
cloud similar to that associated with SBS~0335--052,
 are needed to understand better the triggering of star formation in these 
systems.

\subsection{Star formation threshold}

Star formation in disk galaxies is thought to occur above a critical
threshold of the gas surface density above which clouds will become
gravitationally unstable against collapse (Toomre 1964; Kennicutt
1989). This critical density is proportional to the velocity
dispersion of the gas disk and the epicyclic frequency, which is twice
the circular frequency for regions with solid body rotation.  The
validity of the Kennicutt (1989) HI column density threshold for onset
of star formation has been checked for BCDs (e.g., Taylor et al.\
1994) as well as for LSB galaxies (van der Hulst et
al. 1993). Alternatively, as first mentioned by Skillman (1987),
star formation in dwarf irregulars occurs when the column
densities reach values of order 5 $\times$ 10$^{20}$cm$^{-2}$, with
this threshold increasing with decreasing metallicity (Franco \& Cox
1986). 

The observed peak column densities in the SBS~0335--052 system
coinciding with the optical counterparts are near the upper end of the
range of those reported for HII galaxies by Taylor et al. (1993, 1995)
--- (0.8 -- 11)$\times$ 10$^{20}$cm$^{-2}$ --- when compared to
data on objects mapped with comparable linear resolution, of order
5\,kpc. These values are also in the same range observed for LSB
galaxies, --- (1.0 -- 10.5) $\times$
10$^{20}$cm$^{-2}$ --- objects in which no significant star formation takes 
place, and observed by van der Hulst et al.\ (1993) with a similar linear 
resolution. 

Given our modest spatial resolution, all we can do is make an order of
magnitude estimate of the expected threshold value and compare that to
the observed (beam--smeared) ones. To obtain an estimate we assume
that both peaks fall within the solid body part of the rotation curve,
at a rotational velocity of about 20 \,km\,s$^{-1}$ and a radial
distance of 6 kpc. With a typical gas velocity dispersion $\sigma_{v}$
of 6 \,km\,s$^{-1}$, we obtain for the critical surface density at
the radius of the current SF burst a value equal to only 1.3
$M_{\odot}$ pc$^{-2}$ (equation 6 of Kennicutt 1989 with $\alpha$ = 2/3), 
in comparison with measured values of 8.0
$M_{\odot}$ pc$^{-2}$ and 6.0 $M_{\odot}$ pc$^{-2}$.
With a viewing angle of the HI disk of less than 77$^{\circ}$, the 
gravitational instability condition would be satisfied. This appears to be 
the case since in the simple model 
of a circular flattened HI disk, 
the observed axial ratio would correspond to a 
viewing angle of 70$^{\circ}$.


\subsection{The role of a dark matter halo}

It has been shown (e.g. de Blok et al.\ 1996) that LSB galaxies are
dominated by dark matter over the entire area where HI and stellar
emission is detected. This dark matter dominance is one of the main
factors which allows an LSB galaxy to be stable against various
perturbations (see e.g., the study of the relative fragility of LSB
and high-surface-brightness disks by Mihos et al.\ 1997).  
If in the SBS~0335--052 system
we indeed observe the formation of a first generation of stars in a
low density primordial gas cloud, then its dark matter halo
should have played an especially important role for stabilizing the
gas against various local and global instabilities and preventing it
from gravitationally collapsing and forming stars.  Presumably, a
strong tidal perturbation due either to a close passage of a massive
galaxy or to the interaction with another HI cloud
 can overcome the stabilizing influence of the dark matter halo and
trigger a starburst.

\section{Conclusions}

 From the observational data presented here, and the properties derived
for the HI gas and based on the discussion above we arrive at the
following conclusions:

\begin{enumerate}

\item{The optical dwarf galaxy SBS~0335--052 is associated with a huge HI
cloud with overall size of 66 $\times$  22\,kpc, elongated in the EW
direction. The cloud has a total HI mass of $1.68 \times 10^{9}$
$M_{\odot}$.}   

\item{This HI cloud contains two prominent HI peaks, located nearly
symmetrically to the East and West, and separated by about 22
kpc. They measure 7 $\times$ 4 kpc and contain 0.79 $\times 10^{9}$
$M_{\odot}$ and 0.89 $\times 10^{9}$ $M_{\odot}$, respectively, of
HI.}

\item{The two HI density peaks are identified with two optical
emission--line dwarf galaxies, the eastern one with SBS~0335--052
itself, and the western one with a new, one order of magnitude less
luminous dwarf named SBS~0335--052W. This latter
galaxy shows
very similar properties, including an extremely low metallicity of the
HII region gas, consistent with the amount produced during a first SF
burst.  Radial velocities of these galaxies, as measured from the
emission lines of their HII regions are very close to those of the HI
gas at the positions of their respective peaks.}

\item{The velocity field of the HI cloud is rather disturbed, with
presumably tidal tails protruding from both edges of the cloud. This
suggests that what we are seeing is an interaction, possibly with the nearby 
galaxy NGC 1376. This large spiral galaxy, the dominant member of the
galaxy group LGG 103, with a projected radial velocity of 120\,km\,s$^{-1}$
larger than that of SBS~0335--052, is located at a projected distance
of 150 kpc.
Consideration of the relevant parameters of the SBS~0335--052 system
and NGC~1376 does not contradict the hypothesis that the formation of
the two young dwarfs from the huge low--surface density HI cloud was
triggered by a tidal interaction with NGC~1376.} 

\item{ Alternatively, we are dealing with two almost equal mass HI
clouds, each measuring $\sim 7
\times 4 $\,kpc, each containing an optical counterpart, and showing
solid--body 
rotation with an amplitude of $\sim$ 35-40 \,km\,s$^{-1}$. In this case,
the triggering of the star formation in the two HI clouds is probably 
due to their mutual tidal interaction.   }

\item{Each component seems to require the presence of dark
matter, and the system as a whole is dominated by a dark matter halo. The
total dynamical mass of the system is estimated to be between 
3.8 $\times$ 10$^{9}$ $M_{\odot}$ and 9 $\times$ 10$^{9}$ $M_{\odot}$, or 
$\sim$ 1.9 to  4.5
times larger than the total mass of gas and stars.}

\item{If the viewing angle of the HI
disk is less than 77$^{\circ}$, 
the gas surface densities for both HI peaks are larger than the
critical value determined from the parameters of the rotation curve at
those radii and are consistent with the onset of gravitational
instability and subsequent star formation.}

\end{enumerate}

{\it Acknowledgements.} We thank the partial financial support of 
NATO collaborative research grant 921285 (SAP, TXT), INTAS grant No. 94--2285
(SAP, YII),  NSF grant AST-9616863 (TXT, YII), Russian grant RFBR 
97$-$02$-$16755 (SAP) and  CONACyT grant No.\ 0460P--E (EB).
  SAP is grateful to 
the NRAO staff at Socorro for hospitality and support with the VLA
data reduction, and to Oleg Verkhodanov and Pat Murphy for help with
installing {\sc aips} at the Special Astrophysical Observatory.
The $B$ optical image of the SBS~0335--052 system was kindly 
provided by Polychronis Papaderos. The anonymous referee made useful 
suggestions which helped to improve the presentation of the paper. We have 
made use of the NASA/IPAC Extragalactic Database (NED) which is operated by
 the Jet Propulsion Laboratory, California Institute of Technology, under 
contract with the National Aeronautics and Space Administration. We have 
also used the Digitized Sky Survey which was produced at the Space Telescope 
Science Institute under US Government grant NAG W-2166.

\end{document}